\documentstyle[]{mn}

\newlength{\colwidthf}
\newlength{\hcolwidthf}
\newlength{\colwidth}
\newlength{\hcolwidth}
\font\ninerm=cmr9    \font\sixrm=cmr5
  
\font\nineit=cmti9  
\font\ninesl=cmsl9
\font\ninei=cmmi9    \font\sixi=cmmi5
\skewchar\ninei='177
\font\ninesy=cmsy9  \font\sixsy=cmsy5
\skewchar\ninesy='60
\font\ninebf=cmbx9  \font\sixbf=cmbx5
\font\nineex=cmex10 scaled 833

\font\ninett=cmtt9
\def\adjustlinespace{\baselineskip=\baselineskip}
\def\ninepoint{\textfont0=\ninerm \scriptfont0=\sixrm 
                \def\rm{\fam0\ninerm}\relax
                \textfont1=\ninei \scriptfont1=\sixi 
                \def\mit{\fam1}\def\oldstyle{\fam1\ninei}\relax
                \textfont2=\ninesy \scriptfont2=\sixsy 
                \def\cal{\fam2}\relax
                \textfont3=\nineex \scriptfont3=\nineex 
                \def\it{\fam\itfam\nineit}\relax
                \textfont\itfam=\nineit
                \def\sl{\fam\slfam\ninesl}\relax
                \textfont\slfam=\ninesl
                \def\bf{\fam\bffam\ninebf}\relax
                \textfont\bffam=\ninebf \scriptfont\bffam=\sixbf
                \def\tt{\fam\ttfam\ninett}\relax
                \textfont\ttfam=\ninett
                \setbox\strutbox=\hbox{\vrule
                     hnine7pt depth2pt width0pt}\baselineskip=9pt
                \adjustlinespace
                \rm}
\font\fivebmi=cmmib6
\font\sixbmi=cmmib6     \skewchar\sixbmi='177
\font\ninebmi=cmmib10 at 9pt    \skewchar\ninebmi='177
\newfam\bmifam
\textfont\bmifam=\ninebmi
\scriptfont\bmifam=\sixbmi
\scriptscriptfont\bmifam=\fivebmi

\def\etal{{\em et~al.\ }}
\def\aa #1 #2 {A\&A, #1, #2}
\def\aas #1 #2 {A\&AS, #1, #2}
\def\acm #1 #2 {ACM-Trans Math Software, #1, #2}
\def\ada #1 #2 {Ann Astrophys, #1, #2}
\def\agabstr #1 #2 {Astr Ges Abstr Ser, #1, #2}
\def\aj #1 #2 {AJ, #1, #2}
\def\anach #1 #2 {Astr Nachr, #1, #2}
\def\apj #1 #2 {ApJ, #1, #2}
\def\apjl #1 #2 {ApJL, #1, #2}
\def\apjs #1 #2 {ApJS, #1, #2}
\def\araa #1 #2 {ARAA, #1, #2}
\def\apss #1 #2 {ApSpaceS, #1, #2}
\def\celmech #1 #2 {Cel Mech, #1, #2}
\def\esom #1 #2 {ESO Messenger, #1, #2}
\def\fundcp #1 #2 {FunCosP, #1, #2}
\def\jcp #1 #2 {J Comp Phys, #1, #2}
\def\jfm #1 #2 {J Fluid Mech, #1, #2}
\def\jmp #1 #2 {J Math Phys, #1, #2}
\def\ma #1 #2 {Mitt Astr Ges, #1, #2}
\def\mn #1 #2 {MNRAS, #1, #2}
\def\nat #1 #2 {Nat, #1, #2}
\def\obs #1 #2 {Observatory, #1, #2}
\def\pasj #1 #2 {PASJ, #1, #2}
\def\pasp #1 #2 {PASP, #1, #2}
\def\phyr #1 #2 {PhysRep, #1, #2}
\def\physd #1 #2 {Physica D, #1, #2}
\def\rpp #1 #2 {RepProgPhys, #1, #2}
\def\ssr #1 #2 {Sp Sci Rev, #1, #2}
\def\iau127#1{in de Zeeuw P.T. ed, Structure and Dynamics of 
     Elliptical Galaxies, IAU Symp.~No.~127. Reidel, Dordrecht, p.~#1}

\def\inbook#1#2#3#4#5#6{in: #1%
\if#2-%
\else%
, #2%
\fi%
\if#3-%
\else%
, ed.\ #3%
\fi%
\if#5-%
 {\if#4-%
 \else,%
   (#4)%
 \fi}%
\else%
 {\if#4-%
, (#5)%
\else%
, (#5:#4)%
\fi}%
\fi%
\if#6-%
.%
\else%
, #6.%
\fi%
}

\def\spose#1{\hbox to 0pt{#1\hss}}
\def\lta{\mathrel{\spose{\lower 3pt\hbox{$\mathchar"218$}}
     \raise 2.0pt\hbox{$\mathchar"13C$}}}
\def\gta{\mathrel{\spose{\lower 3pt\hbox{$\mathchar"218$}}
     \raise 2.0pt\hbox{$\mathchar"13E$}}}

\def\gt{\! > \!}
\def\lt{\! < \!}

\def\=#1{\overline{#1}}

\def\lvplot{($l,v$) diagram}
\def\lvplots{\lvplot s}

\def\phibar{\varphi_{\rm bar}}

\def\deg{^\circ}             

\def\mum{\mu{\rm m}}         

\def\kms{{\rm\,km\,s^{-1}}}

\def\pc{{\rm\,pc}}
\def\kpc{{\rm\,kpc}}

\def\Gyr{{\rm\,Gyr}}

\def\lvplot{($l,v$) diagram}
\def\lvplots{\lvplot s}
\def\tvc{{\rm TVC}}
\def\tvcs{{\tvc s}}

\def\phibar{\varphi_{\rm bar}}

\def\kpc{\,{\rm kpc}}
\def\Gyr{\rm Gyr}
\def\kms{{\rm\,km\,s^{-1}}}

\def\etal{{\rm et al.}}

\def\thalf{{\textstyle{1\over2}}}

\title{Gas Dynamics in the Milky Way: Second Pattern Speed and
  Large--Scale Morphology}

\author
[Nicolai Bissantz, Peter Englmaier, and Ortwin Gerhard]
{Nicolai~Bissantz$^{1}$\thanks{Present address: Institut f\"ur 
Mathematische Stochastik der Universit\"at G\"ottingen, 
Lotzestr. 13, 37083 G\"ottingen, Germany}, 
Peter~Englmaier$^2$, and Ortwin~Gerhard$^1$ \\
$^1$Astronomisches Institut, Universit\"at Basel,
Venusstrasse 7, CH-4102 Binningen, Switzerland \\
$^2$Max-Planck Institut f\"{u}r extraterrestrische Physik, 
Garching, Germany }

\begin{document}

\maketitle

\begin{abstract}
  We present new gas flow models for the Milky Way inside the solar
  circle. We use SPH simulations in gravitational potentials
  determined from the NIR luminosity distribution of the bulge and
  disk, assuming constant NIR mass-to-light ratio, with an outer halo
  added in some cases. The luminosity models are based on the {\it
    COBE/DIRBE} maps and on clump giant star counts in several bulge
  fields, and include a spiral arm model for the disk.
  
  Gas flows in models which include massive spiral arms clearly match
  the observed $^{12}$CO \lvplot\ better than if the potential does
  not include spiral structure. Furthermore, models in which the
  luminous mass distribution and the gravitational potential of the
  Milky Way have four spiral arms are better fits to the observed
  \lvplot\ than two-armed models.
  
  Besides single pattern speed models we investigate models with
  separate pattern speeds for the bar and spiral arms.  The most
  important difference is that in the latter case the gas spiral arms
  go through the bar corotation region, keeping the gas aligned with
  the arms there. In the ($l,v$) plot this results in characteristic
  regions which appear to be nearly void of gas.  In single pattern
  speed models these regions are filled with gas because the spiral
  arms dissolve in the bar corotation region.
  
  Comparing with the $^{12}$CO data we find evidence for separate
  pattern speeds in the Milky Way.  From a series of models the
  preferred range for the bar pattern speed is $\Omega_p=60\pm 5
  \Gyr^{-1}$, corresponding to corotation at $3.4\pm0.3 \kpc$.
  The spiral pattern speed is less well constrained, but our preferred
  value is $\Omega_{sp}\approx 20 \Gyr^{-1}$. A further series of gas
  models is computed for different bar angles, using separately
  determined luminosity models and gravitational potentials in each
  case. We find acceptable gas models for $20\deg\lta\phibar
  \lta25\deg$. The model with ($\phibar=20\deg$,
  $\Omega_p=60\Gyr^{-1}$, $\Omega_{sp}=20 \Gyr^{-1}$) gives an
  excellent fit to the spiral arm ridges in the observed ($l,v$) plot.
\end{abstract}


\section{Introduction}
\label{s1}
Observations of cold gas in the Milky Way [MW] have contributed
substantially to our understanding of MW structure. No other tracer is
observed in as large a part of the MW as are gas clouds.
Longitude-velocity [lv] diagrams (Hartmann \& Burton \shortcite{HB97},
Dame et al.\ \shortcite{Dame01}) show the distribution of gas
velocities as a function of galactic longitude $l$, integrated over
some range in latitude $b$.  By observing the MW in different spectral
lines, this gas can be traced at substantially different densities.
The largest absolute velocity as a function of $l$ defines the
terminal velocity curve (\tvc). In an axisymmetric galaxy, the gas at
these velocities is found at the ``tangent point'' where the line of
sight is tangential to a circle around the Galactic centre. From this
the rotation curve can be determined. However, due to the bar and
spiral perturbations in the MW potential, the gas has substantial
non-circular velocities, which are most evident in the central
$10\deg-20\deg$, due to the bar, but also as ``bumps'' in the \tvc\ 
where spiral arm tangents perturb the gas flow by $\sim 10$-$20 \kms$.
At sub-\tvc\ velocities, crowding in both position and in velocity
produces ridge-like structures in the \lvplot. The Galactic spiral
arms are visible as straight or curved such ridges.

A number of attempts have been made to model these observations.  One
group is formed by analytic models of spiral structure.  The first
exhaustive analytical formulation of a spiral arm theory was developed
by Lin \& Shu \cite{Lin64} and applied to the MW by Lin et al.\ 
\cite{Lin69}. They proposed a two-armed model with pitch angle
$-6\deg$ and a pattern speed for the spiral structure $\Omega_{sp}
\approx 13.5 {\rm km/s/kpc}$. Amaral \& L\'epine \cite{AmLep97} fitted
the rotation curve of the MW to an analytic mass model and found a
self-consistent solution with a combined two- and four-armed spiral
structure. In L\'epine et al.\ \cite{Lep01} they extended this model
to allow for a phase difference between the two-armed and the
four-armed spiral pattern.

The second group, numerical simulations of the Galactic gas flow, also
have a long tradition.  A recent example is the smoothed particles
hydrodynamics (SPH) models of Fux \cite{Fux99}, who evolved a gas disk
inside a self-consistent $N$-body model scaled to the {\it COBE/DIRBE}
K-band map of the MW and the radial velocity dispersion of M giants in
Baade's window. The resulting gas flow was transient, but at specific
times resembled closely a number of observed arms and clumps in the
bar region.  Weiner \& Sellwood \cite{Weiner99} compared predictions
from fluid dynamical simulations for analytic mass densities with the
observed outer velocity contours of the HI $(l,v)$ diagram to
constrain the pattern speed and bar angle.  Englmaier \& Gerhard
\shortcite{EG99}  [hereafter, Paper I] computed gas flows in the
gravitational potential of the NIR luminosity distribution of Binney,
Gerhard \& Spergel \cite{BGS97}, assuming constant NIR mass-to-light
ratio ($M/L$). Their best SPH gas flow models reproduced
quantitatively a number of observed gas flow features, including the
positions of the five main spiral arm tangents at $|l|\leq 60\deg$ and
much of the terminal velocity curve.

An important feature of all these gas flow models is the pattern speed
of the non-axisymmetric component. Englmaier \& Gerhard found a best
pattern speed for the bar $\Omega_p\approx 60 {\rm km/s/kpc}$.  Weiner
\& Sellwood derived a bar pattern speed $\approx 42 {\rm km/s/kpc}$,
whereas
Fux determined $\approx 50 {\rm km/s/kpc}$ from his models.  Dehnen
\cite{Dehn00} used resonant features in the Hipparcos stellar velocity
distribution to argue that the Sun is located just outside the OLR of the
exciting quadrupole perturbation, giving a pattern speed of $\approx
51 {\rm km/s/kpc}$ for solar constants $R_0=8\kpc$ and $v_0=220\kms$.
For their {\it spiral structure} model, Amaral \& L\'epine
\cite{AmLep97} and L\'epine et al.\ \cite{Lep01} found
$\Omega_{sp}\approx 20\, \Gyr^{-1}-35\, \Gyr^{-1}$.  Fern\'andez \etal\ 
\cite{Fer01} obtained a somewhat higher $\Omega_{sp}\approx 30 {\rm
  km/s/kpc}$ from Hipparcos data for OB stars and Cepheids.
Debattista et al.\ \cite{Deba02} used the
Tremaine-Weinberg method on a sample of intermediate age to 8 Gyr old
OH/IR-stars in the inner Galactic disk.  They found a pattern speed of
$59 \pm 5 \pm 10 $ (systematic) km/s/kpc, which may be driven by the
bar in the center of the MW.

Thus the bar and spiral arms in the MW may not rotate with the same
pattern speed.  For a fast bar, a single $\Omega_p$ would imply that
the spiral arms are entirely {\sl outside} their corotation radius.
Observations of external galaxies (see, e.g., the Hubble Atlas of
Galaxies, Sandage \shortcite{Sandage61}) suggest that galaxies exist
with dust lanes on the inner (concave) edges of their spiral arms. For
a trailing spiral pattern, these arms would be {\it inside} their
corotation radius.  A lower pattern speed for the spiral structure
than for the bar would remove this discrepancy.  Indeed, Sellwood \&
Sparke \cite{Sell88} showed evidence for multiple pattern speeds in
their $N$-body simulations.  Rautiainen \& Salo \cite{Rau99} analysed
two-dimensional $N$-body simulations, some of them with a massless,
dissipative gas component added.  They confirmed the possibility of
multiple pattern speeds in self-consistent N-body models of 
barred galaxies, and found a number of
possible configurations. These included models with corotating bar and
spirals, as well as models with different pattern speeds. Some of the
models in the latter group show evidence for a non-linear
mode-coupling (Tagger \etal\  \shortcite{Tagger87}) between bar and
spiral pattern, but others show no such evidence.  In some of their
models there exist separate inner spirals corotating with the bar, and
outer spirals which rotate with their own, lower pattern speed.  It is
therefore tempting to analyse gas flow models of the MW with multiple
pattern speeds.

What is the morphology of the MW spiral arms?  Most authors infer four
spiral arms from tracers which directly or indirectly measure the gas
density, such as molecular clouds, HII regions, pulsars, and the
galactic magnetic field (Georgelin \& Georgelin \shortcite{gg76},
Sanders et al.\ \shortcite{Sanders85}, Caswell \& Haynes
\shortcite{CasHay87}, Grabelsky \& et al.\ \shortcite{Grab88}, Taylor
\& Cordes \shortcite{TayCor93}, Vall\'ee \shortcite{Val95}; however,
Bash \cite{Bash81} infers a two-armed pattern from the same HII-data
as used by Georgelin \& Georgelin). The problem is that all spiral arm
parameters other than the tangent point directions (e.g., Paper I)
require distance information. It is also not clear whether all spiral
arms seen in the MW gas are present in the old disk. Ortiz \& L\'epine
\cite{Ort93} constructed a four-armed model which reproduces their
star counts in the near infrared.  Drimmel \cite{Drimmel00} preferred
a two-armed structure from {\it COBE/DIRBE} K-band data, but a
four-armed structure for the dust distribution seen in the $240\mum$
data.  Drimmel \& Spergel \cite{Drimmel01} project a luminosity model
through the $240\mum$ dust model, to compare with the NIR J and K band
{\it COBE/DIRBE} data.  Their best model for the stellar distribution
is four-armed, but dominated by two spiral arms.  Drimmel \& Spergel
conclude that, if there are four arms in the K-band luminosity
distribution, the Sag-Car arm is of reduced strength (by a factor of
$2.5$).

In this paper we investigate the dynamical effects of the Galactic bar
and spiral arms on the gas flow in the Milky Way.  We investigate the
possibility of different pattern speeds for bar and spiral arms, and
the consequences this would have on the observed \lvplots.  Our mass
models for the inner Galaxy are based on the NIR luminosity density
models of Bissantz \& Gerhard \cite{BissGerh02} 
[hereafter, Paper II] which include spiral
structure.  We use SPH simulations to determine the gas flow in the
MW, extending the work presented in Paper I, where eightfold symmetric
mass models were used.  This paper is organised as follows.  In
Section \ref{s2} we describe the luminosity models, methods, and
observational data used in this work.  In Section \ref{s3} we describe
our best gas model for the observed $^{12}$CO \lvplot. Then we compare
models with different pattern speeds (Section \ref{s4}), bar angles
and spiral arm morphology (Section \ref{s5}) to constrain these
parameters, and finally give our conclusions in Section \ref{s7}.

\section{Gas dynamical model}
\label{s2}

\subsection{Mass model of the Milky Way}
\label{s21}
\label{PMW}
Here we first describe the adopted model for the distribution of
luminous mass in the MW. The model is based on the dust-corrected
near-infrared maps of Spergel et al.\ \cite{Spergel96}, which they
obtained from {\it COBE/DIRBE} data using a three-dimensional dust
model derived from $240\mu m$ observations.  From their L-band map we
obtained a non-parametric luminosity distribution using the procedure
described in Bissantz \& Gerhard \cite{BissGerh02}.  
Because the non-parametric model only covers the central $5\kpc$
part of the MW, we used a parametric best-fit model of the same L-band
data to extend the model to larger radii. To convert the luminosity
model to a model for the luminous mass density we assumed constant
L-band mass-to-light ratio.

The model building procedure is described in detail in Paper II. Here
we only summarise the most important points. Bissantz \& Gerhard
estimated the luminosity model iteratively from the L-band data by
maximising a penalised likelihood function.  The penalty terms
encourage eightfold-symmetry with respect to the three main planes of
the bar, as well as smoothness and a prescribed spiral structure model
in the disk. The spiral structure term is based on an approximate
model for the MW spiral arms from Ortiz \& L\'epine \cite{Ort93}.  The
best models resulting from this approach reproduced the dust-corrected
{\it COBE/DIRBE} maps with an RMS accuracy of 0.07 mag.

To define the eightfold symmetry penalty term we had to specify the
position of the Sun in the MW. Bissantz \& Gerhard set the distance of
the Sun to the galactic centre to $R_0=8\kpc$, and the vertical
distance from the galactic plane to $z_0=14\pc$. These values will be
used throughout the present paper as well.  The third parameter is the
bar angle. Paper II compared photometric models for bar angles
$10\deg\leq\phibar\leq 44\deg$, and concluded that the best models had
bar angles $20\deg\leq\phibar\leq 25\deg$.

Paper II also showed that models which include spiral structure are
better than models without spiral structure. In particular, because
some spiral arm tangent points are evident in the L-band data, such models
give a better description of the nearby disk.  Also, the inclusion of
spiral arms makes a model appear broader on the sky. For given data,
this allows the bulge/bar to be more elongated in models with spiral
structure. The larger {\sl bulge} elongation in these models makes it
possible to reproduce the asymmetries seen in the apparent magnitude
distributions of clump giant stars in several bulge fields (Stanek et
al.\ \shortcite{Stanek94}, \shortcite{Stanek97}), for
$15\deg\lta\phibar\lta 30\deg$.  The shape of the bulge/bar in the
model with $\phibar=20\deg$ is about $10:3-4:3-4$, and its length is
approximately $3.5\kpc$.

\subsection{Gravitational potential}
\label{s22}
The gravitational force field and potential generated by the
distribution of luminous mass were calculated using the multipole
expansion method described in Paper I, modified to include phase terms
in order to allow for the spiral arm components. The multipole
expansion method was used because (i) it allows an independent
treatment of disk, bar, and spiral arm components, (ii) it works for
an arbitrary density distribution, and (iii) it does not require
detailed boundary conditions as do other methods, since the
integrations can easily be extended to infinite radii.  The expansion
was computed to maximum spherical harmonic orders $l_{max}=8$ and
$m_{max}=8$.  Odd orders were neglected, assuming point-symmetry with
respect to the centre.  Naively, one could interpret the $m=2$
component as due to the bar mode and all higher $m$ modes as due to
spiral arm modes. However, some of the $m=2$ component beyond the
bar's corotation radius is due to an $m=2$ spiral arm mode in the
luminosity model (see also Amaral \& L\'epine \shortcite{AmLep97}).

Since the spatial resolution of the mass models obtained in Paper II
is limited by that of the dust-corrected maps of Spergel et al., the
central cusp in the MW's density distribution and potential is
incorrectly represented in our multipole expansion model. We attempted
to correct for this by modifying the multipole coefficient functions
in the centre as in Paper I, replacing the central mass distribution
by a power law $\rho^{-1.8}$.

Optionally we add an analytical halo potential
\[
  \phi_{\rm Halo} = {\thalf} V_{\rm inf}^2 \ln{(r^2+a^2)},
\]  
where $V_{\rm inf}$ is the circular velocity at infinity and $a$ is the
core radius. In this paper we use $V_{\rm inf}=220 {\rm km/s}$ and 
determine the core radius such that the model in the potential with
halo best fits the \tvc\  near $|l|=90\deg$.

\begin{figure}
  \caption[]{Rotation curve of the standard mass model for bar angle
$20\deg$. The velocities have been scaled with the factor $\xi$ determined
in Section \ref{s3}, fitting the observed terminal velocities
by the SPH model in $10\deg<|l|<50\deg$.}
  \label{rotcurve}
\end{figure}    

Figure \ref{rotcurve} shows the rotation curve derived from the
$m$=0--component of the reference luminosity density model of Paper II
(for bar angle $20\deg$, with the parametric extension). For brevity,
this mass density is called "standard mass model" hereafter, and the
associated gravitational potential "standard potential".  Its rotation
curve no longer shows a strong bump in the inner Galaxy's rotation
curve as for the models of Paper I (Fig.~4 there). This is because the
new luminosity model reproduces light in the MW's spiral arms
significantly better, as compared to the density maxima $\sim 3\kpc$
down the minor axis of the bar in the earlier models. We note that the
spiral arms in the luminosity models used here arbitrarily start at a
galactocentric radius of $3.5\kpc$.  This generates a slight
distortion of the rotation curve around this radius.

\begin{figure}
  \caption[]{Quadrupole and octopole terms of the standard  
potential, separated into the part generated by the triaxial bar/bulge in the 
distribution of luminous mass (full line: $m\!=\!2$, dotted line: $m\!=\!4$),
and the part caused by the spiral arms (short dashed: $m\!=\!2$, long dashed: 
$m\!=\!4$). All multipoles are normalised by the value of the 
monopole term at the respective Galactic radius. }
  \label{potbild}
\end{figure}    

Figure \ref{potbild} shows the quadrupole and octopole terms of the
potential, separated into the parts generated by the bar and the parts
generated by the spiral arms.  All multipoles are normalised by the
value of the monopole term at the respective Galactic radius.  Note
that in this model the bar plays a major role for the non-axisymmetric
forces, in particular the quadrupole moment, even beyond where it ends
in the mass density.

The multipole representation of the potential is used in the orbital
analysis and in the two-dimensional SPH hydrodynamics code, as
described in Paper I.  The bar and spiral components are both given a
constant pattern speed.  In most models described below, the bar and
spiral arm patterns rotate with different pattern speeds, implying
that the mass distribution and potential undergo periodic
oscillations; in some models these pattern speeds are equal and
the mass distribution is constant in the rotating frame.

\begin{figure}
  \caption[]{Resonance diagram for the standard mass model, assuming
the same velocity scaling as in Fig.~\ref{rotcurve}. }
  \label{figres}
\end{figure}    

Figure \ref{figres} shows a resonance diagram for the standard mass
model with the same scaling as in Figure~\ref{rotcurve}.  (The scaling
constant is different by $\lta 1.5\%$ for models {\it 40} and {\it
  60}, which have significantly different spiral pattern speeds from
our standard model.)  We will see below that the preferred bar pattern
speed is $\Omega_p=60 \Gyr^{-1}$. Observe from Fig.~\ref{figres} that
in this case the corotation radius of the bar nearly coincides with
the inner ultra-harmonic resonance of the spiral structure, if this
has pattern speed $\Omega_{sp}=40 \Gyr^{-1}$, and with its inner
Lindblad resonance, if $\Omega_{sp}=20 \Gyr^{-1}$. Models with these
values for the spiral arm pattern speed in conjunction with
$\Omega_p=60 \Gyr^{-1}$ are particularly interesting to analyse
(Tagger \etal\ \shortcite{Tagger87}), and are discussed below.

\subsection{Hydrodynamical method}
\label{s23}
We use the two-dimensional smoothed particles hydrodynamics [SPH] code
described by Englmaier \& Gerhard \cite{EG97}. The code solves
Euler's equation for an isothermal gas with effective sound speed $c_s$:
\begin{eqnarray}
\frac{\partial {\bf v}}{\partial t}+{\bf (v\cdot\nabla)v}
= -c_s^2 \frac{{\bf\nabla} \rho}{\rho} - {\bf\nabla}\Phi.
\end{eqnarray}
This approach is based on a result of Cowie \cite{Cowie80}.
He showed that an isothermal single fluid description  
crudely approximates the dynamics of the ISM. However, here
the isothermal sound speed is not the thermal sound speed,
but an effective sound speed representing the RMS random
velocity of the cloud ensemble. 

The SPH method has the advantage of allowing for a spatially adaptive
resolution length. This is achieved by adjusting the smoothing length
$h$ of a particle everywhere such that the number of particles that
overlap a given particle is approximately constant. Fluid quantities
are approximated by averaging over neighbouring
particles. Furthermore, the SPH scheme includes an artificial
viscosity to allow for shocks in the simulated gas flow. For further
discussion of the method see Englmaier \& Gerhard \cite{EG97}, Paper
I, and Steinmetz \& M\"uller \cite{Stein93}.

Our SPH models contain $50000-60000$ particles, except when indicated
otherwise.  The initial surface density of the models is taken to be
constant inside $8\kpc$ galactocentric radius.

\begin{figure*}
  \caption[]{Central part of the $^{12}$CO-observations of Dame et
al.\ \cite{Dame01}.  White lines sketch spiral arm ridges in the data.}
  \label{dameco}
\end{figure*}

\subsection{Comparison with observational data}
\label{s24}

Our main tools to compare our gas flow models with observations are
the terminal velocity curve (\tvc ) and the \lvplot, which shows
the radial velocities of gas clouds as function of galactic longitude.
Throughout this paper observed velocities are given with respect to
the local standard of rest (LSR).  For the distance of the Sun from
the galactic centre we assume $R_0=8\kpc$.  The LSR circular velocity
is assumed to be $v_0=220{\rm km/s}$, consistent with $R_0=8\kpc$
(Feast \& Whitelock \shortcite{FW97}, Reid et al.\ \shortcite{reid99}, Backer \&
Sramek \shortcite{backer99}).  Note that a $10\%$ change in $v_0$ is not
critical, amounting to radial velocity variations of only $\sim
10\kms$ in the central $l=45\deg$ (cf.\ Paper I).

\lvplots\ for our gas flow models are constructed as follows.  The LSR
observer is specified by the galactocentric radius $R_0$, the LSR
circular velocity $V_0$, and by the angle $\phibar$ relative to the
bar.  We first project all particle velocities $(v_x,v_y)$ onto the
line-of-sight vector $\vec{e}_p$ from the LSR observer to the
particle, subtracting the component of $V_0$ in the direction of the
particle:
\begin{equation}
v_{r}=\vec{e}_p \cdot (v_x,v_y) - v_0 \sin(l).
\end{equation}
We then construct a two-dimensional binned histogram of the
particle distribution in the $l-v$-plane, with bin size 
$\approx 0.23^{\deg}\times 0.7{\rm km/s}$. Finally, we convert
this histogram into a greyscale plot, using a lower surface
density cutoff $C_{\rm lv}=0.5-1\%$ to enhance the contrast. 
$C_{\rm lv}$ varies between different models, and is selected so
as to optimize the visibility of the spiral arm ridges 
and the terminal velocity envelope.

We compare a model with observations in a two step process.  In the
first step, we focus on the terminal velocities, comparing the model
\tvc\ with an observed \tvc\ composed from the following data: HI
velocities from Burton \& Liszt \cite{BurtonLiszt93}; HI velocities
from Fich et al.\ \cite{Fich89}, based on data from Westerhout
\cite{Westh57}; unpublished $140$-ft single dish HI velocities,
kindly provided by Dr. B. Burton; northern $^{12}$CO-velocities from
Clemens \cite{Clemens85}, including error bars; and southern
$^{12}$CO-velocities from Alvarez \etal\ \cite{Alvarez90}.  The
Clemens \cite{Clemens85} data were corrected for internal dispersion
(Paper I).  The observed velocities, corrected by the respective
authors or Paper I to the pre-Hipparcos LSR frame, in which the Sun
was assumed to move with approximately $u_\odot=-10\kms$ inwards and
$v_\odot=15\kms$ in the forward direction of Galactic rotation, are
here transformed to the Hipparcos LSR frame ($u_\odot=-10\kms$,
$v_\odot=5\kms$, Dehnen \& Binney \shortcite{DehBin98}), by
subtracting $10\sin l \kms$.

A free parameter of our models is the mass-to-light ratio. From the
comparison of the model \tvc\ with the observed \tvc\ we determine the
best-fit scaling factor $\xi$ for the model velocities.  This
parameter $\xi$, which determines the mass scale of the model,
varies by $\sim 5-10\%$ between all our models. We only use
the \tvc\ data for $10\deg<|l|<50\deg$ to determine $\xi$, because
near the centre the resolution of our models is insufficient, and
because for $|l|\gta 50\deg$ the halo contributes significantly (see
below).  In the fitting we take special care of the location of
``bumps'' in the \tvc, because these indicate spiral arm tangents.
This procedure assumes that the NIR disk and bulge are responsible for
all of the observed velocities in the central parts of the MW.

In most models the gravitational potential is time dependent, because
the pattern speeds of bar and spirals are different. In this case we
select a ``best'' snapshot, corresponding to a specific phase and
evolutionary age.  The value of $\xi$ generally depends slightly on
both the model and the evolutionary age of the snapshot.

\begin{figure}
  \caption[]{The distribution of gas in the standard gas model at 
    evolutionary age $0.32 \Gyr$. Note that the initial particle
    distribution in the simulation ends at $R= 8\kpc$, producing the
    artifical outer cutoff of the particle disk in the plot. The
    position of the Sun at $(x,y)\approx(7.5\kpc,2.7\kpc)$ is shown by
    the $\odot$ symbol.}
  \label{fig31a}
\end{figure}    

\begin{figure}
  \caption[]{The terminal velocity curve (\tvc) of the standard gas model
    at evolutionary time $0.32 \Gyr$ (top and bottom curves), and of
    model {\it halo} also at time $0.32 \Gyr$ (middle curves),
    compared to the HI and CO data. Model velocities have been scaled
    by a model-dependent factor $\xi$ to best fit the observed
    terminal velocities for $10\deg<|l|<50\deg$. The observed
    velocities are corrected to the Hipparcos LSR frame as described
    in \S\ref{s24}.  The \tvc s of model {\it halo} are offset by $60
    {\rm km/s}$ for better readability of the diagram.}
  \label{fig31e}
\end{figure}    

In the second step we compare the model \lvplot\ with the observed
$^{12}$CO \lvplot\ of Dame et al.\ \cite{Dame01}
(Figure~\ref{dameco}), using the scaling $\xi$ from the \tvc.
Important features in the observed \lvplot\ are the ridges of emission
which indicate the location of spiral arms. The white lines drawn in
Fig. \ref{dameco} reproduce approximately the locations of these
spiral arm ridges. These lines are then transformed to the Hipparcos
LSR [Figure~\ref{dameco} was constructed assuming a solar motion of
$\vert\bmath{v}_\odot\vert=20\kms$ towards
$(l,b)=(56.2^\circ,22.8^\circ)$), and are then overplotted on most
model \lvplots. These observed ridges should be reproduced by a good
model of the MW gas flow.  The reverse need not always be true,
however, because the visibility of a spiral arm ridge in the data may
depend on the radial distribution of gas and the geometry of the arm
with respect to the line-of-sight.

In the model \lvplots\ we also show molecular cloud observations of
Dame et al.\ \cite{Dame86} and Bronfman \cite{Bronfman89} [symbol
``x'' in the plots], and HII region observations of Georgelin \&
Georgelin \cite{gg76}, Downes et al.\ \cite{Downes80}, and Caswell \&
Haynes \cite{CasHay87} [symbol ``+'']. Most of our models do not
contain a halo potential, and therefore underestimate the velocities
for $|l|\gta 40-50\deg$. Thus we omit the molecular cloud and HII
observations in these models for $|l|>40\deg$.  For the sake of
clarity we have also left out clouds with less than
$10^{5.5}M_{\odot}$ from the Bronfman et al.\ data, as well as clouds
in the smallest brightness bin from the Georgelin \& Georgelin sample.
For a more detailed discussion of these observations see Paper I.

\section{Best-fit model for the Milky Way}
\label{s3}

We have investigated a number of gas flow models in the COBE
potentials of Section~\ref{s2}, for different pattern speeds, bar
angles, and stellar spiral arm morphologies. The analysis of these
model sequences is deferred to Section~\ref{s4}. Here we begin with a
description of our best model for the gas dynamics in the MW.  This
best-fit model (hereafter called "standard model") is based on the
standard four-armed $\phibar=20\deg$ luminosity model (Bissantz \&
Gerhard \cite{BissGerh02} -- Paper II), as described in Section 2, it
is point-symmetric, the bar pattern speed is $\Omega_p=60 \Gyr^{-1}$,
implying corotation at $R_{\rm cr}\approx 3.4\kpc$, and the spiral arm
pattern speed is $\Omega_{sp}=20 \Gyr^{-1}$.

The gas distribution of the model is shown in Figure~\ref{fig31a}. In
this plot the Sun is at $(x,y)=(7.5,2.7)\kpc$, with $R_0=8\kpc$. Both
inside and outside corotation there exist four spiral arms which are
connected in a complicated way through the corotation region of the
bar.  Outside corotation the spiral pattern in the gas response
consists of a pair of strong arms and a pair of weaker arms.

\begin{figure*}
  \caption[]{\lvplot\ for the standard gas model at evolutionary age 
    $0.32 \Gyr$, for LSR-velocity $V_0= 220 {\rm km/s}$. The model
    velocities are scaled with the factor $\xi$ determined from
    Fig.~\ref{fig31e}.}
  \label{fig31b}
\end{figure*}   

\begin{figure*}
  \caption[]{Grey-scale $lv$-plot for the standard gas model with
    a dark halo component included in the potential (model {\it
      halo}), at evolutionary age $0.32 \Gyr$. Particles in regions of
    low gas surface density ($<1\%$ of the maximum surface density)
    are suppressed to enhance the contrast, mimicking also the bias of
    the observed distribution of molecular gas and HII regions towards
    higher densities.  Model velocities are scaled such as to fit best
    the observed terminal velocities in $10\deg\leq|l|\leq50\deg$ (see
    Fig.~\ref{fig31e}); the LSR-velocity is $220 {\rm km/s}$.  For
    comparison with observations, the spiral arm ridge lines from
    Fig.~\ref{dameco} are overplotted, as are the data for giant
    molecular clouds from Dame \etal\ \cite{Dame86} and Bronfman
    \cite{Bronfman89} (``x'' symbols), and for HII regions from
    Georgelin \& Georgelin \cite{gg76}, Downes et al.\ 
    \cite{Downes80}, and Caswell \& Haynes \cite{CasHay87} (``+''
    symbols); see \S\ref{s24}.  The short vertical lines mark the
    observed spiral arm tangent directions from Englmaier \& Gerhard
    \shortcite{EG99}.  }
  \label{fig31g}
\end{figure*}   

To compare the model to the gas observations, we scale it to the
observed terminal velocity curve (\tvc ) in the longitude range
$10\deg\leq|l|\leq50\deg$, using an LSR circular velocity
$V_0=220\kms$. With this scaling the CO \tvc\ is
reproduced well by the model (Figure~\ref{fig31e}), including most
distinct ``bumps'' in the observations except that at $l\approx
50\deg$.  Vice-versa, the model \tvc\ shows an extra bump at $l\approx
-15\deg$ which is not seen in the data. Presumably the potential in
this bar--disk transition region is not accurately modelled -- there
is little information in the NIR data on the mass distribution in this
region. Overall, however, the fall-off with longitude and even the
detailed form of the inner disk Galactic \tvc\ are represented well by
the model.

In the bulge region, for $|l|\lta 10\deg$, the model velocities are
limited by the resolution of the hydrodynamic simulation. We thus do
not expect to fit the observed large terminal velocities there, but
several other effects may play a role as well. See \S\ref{s31} below
and Englmaier \& Gerhard \shortcite{EG99}  for a
more detailed discussion.  Because the gas particles in the numerical
simulations flow inwards, however, the velocity structure in the inner
$1-1.5\kpc$ is not of significant relevance for the gas flow and,
particularly, for the shock structure in the main spiral arms well
outside this region.  Only the velocities of the pair of inner arms
passing the minor axis of the bar laterally at $\approx 1\kpc$ could
be somewhat affected. Since the main aim here is to investigate the
{\it large-scale} gas flow and spiral arm morphology, we have
therefore not attempted a detailed fit to the inner bulge terminal
velocities.

The standard model does not contain a dark matter halo.  For the
assumed LSR velocity of $V_0=220\kms$, its \tvc\ falls below the
observed \tvc\ at $|l|\gta 40\deg -50\deg$.  At $|l|=90\deg$ the
deficit in the model's terminal velocity is $\approx25{\rm km/s}$.
This can be corrected by a adding a halo potential that generates a
rotation velocity $v_{DM}\approx120 {\rm km/s}$ near the Sun.  We thus
constructed a new gas model in a potential which includes a suitable
quasi-isothermal dark matter halo, which has circular velocity at
infinity $V_{\rm inf}=220 {\rm km/s}$ and core radius $a=10.7\kpc$
(there is substantial freedom in these parameter values).  The \tvc\ 
of this model {\it halo} is also shown in Fig.~\ref{fig31e}; its
best-fitting scaling factor $\xi$ and mass-to-light ratio is only
slightly different from that of the standard model. Model {\it halo}
fits the shape of the observed \tvc\ out to $R_0$, but is not quite as
good a match to the bumps in the \tvc\ as the standard model. This
suggests that the assumed halo model is oversimplified.  Note the fact
that the constant M/L standard model, a maximum disk model by
construction, reproduces the observed terminal velocities inside
galactocentric radius $R\approx 5\kpc$ and still accounts for most of
the circular velocity near the Sun.

The \lvplot\ for the standard model is shown in Figure~\ref{fig31b}.
The dense ridges in this diagram show the locations of the spiral
arms. Figure 11 of Paper II shows the
correspondence between the locations of spiral arms in the Galactic
plane and in the \lvplot. The model spiral arm ridges in
Fig.~\ref{fig31b} generally coincide very well with the observed
spiral arm ridges. Less good is the correspondence for the 3kpc-arm
(again in the bar--disk transition region), which is at too small
negative velocities in the model compared to the observations for
$l\gta -5\deg$.

For model {\it halo} the lv-plot is shown in Figure~\ref{fig31g}, at
enhanced contrast to emphasize the spiral arms (see \S\ref{s24}).  
Compared to the
standard model \lvplot\ (without halo), significant differences are in
the outer spiral tangents near $l\approx\pm50\deg$ which are relocated
by a few degrees, and in the terminal velocities at $\vert
l\vert\gta\pm50\deg$. This is a general result: For a number of models
discussed later in this paper we have added a suitable halo potential
and computed a new gas model, sometimes additionally changing the
extent of the initial gas disk in the simulation from the standard
$8\kpc$ to $10\kpc$. In all these cases the only significant change
has been that the outer tangent points moved outwards in longitude by
$|\Delta l|\lta5\deg$. Thus we do not include a halo potential in
the remaining models discussed below.

Also shown in Fig.~\ref{fig31g} are the observed spiral arm ridge
lines, tangent points, and tracers from \S\ref{s24}. The comparison
with the grey scale plot for the model shows that model {\it halo}
gives a very good description of the gas kinematics in the disk
outside the bar.  For many features deviations are less than $\sim
10\kms$.

We end this section with showing in Figure~\ref{fig31c} a map of the
radial velocities of gas clouds with respect to the LSR for model {\it
  halo}. This map allows one to assess the likely errors made in
determining kinematic distances from cloud radial velocities by
assuming a circular orbit model. 

\begin{figure}
  \caption[]{Contours of constant radial velocity for gas clouds 
    in model {\it halo}, as seen by an observer moving with the
    velocity of the LSR. Contours are spaced by $10 {\rm km/s}$.
    Dashed contours indicate negative radial velocities, full contours
    positive radial velocities. Ticks along the full line give
    distances from the LSR (right end), in kpc along the line-of-sight
    through the Galactic Center. Longitudes as seen from the LSR are
    indicated on the margin of the plot.}
  \label{fig31c}
\end{figure}

\begin{figure}
  \caption[]{A sample of closed x$_1$ and x$_2$ orbits in the bar frame,
    in the $\phibar=20\deg$ potential of our standard gas model
    with pattern speeds $\Omega_p=60 \Gyr^{-1},
    \Omega_{sp}=20 \Gyr^{-1}$. Note the convergence of the outer 
    x$_2$-orbits on the major axis of the bar.}
  \label{fig35b}
\end{figure}   

\begin{figure}
  \caption[]{The same orbits as shown in Fig.~\ref{fig35b}, now displayed in 
    an \lvplot, using $\phibar=20\deg$ also for the projection.  The
    orbit with the highest peak velocity, at $l\simeq 2\deg$, is the
    cusped orbit. Inside this cusped orbit two x$_1$-orbits with
    self-intersecting loops are plotted (cf.~Fig \ref{fig35b}).
    Slight oscillations in the $(l,v)$ traces of the outer x$_1$
    orbits betray the time-dependence of the potential. All orbit
    velocities are scaled by the same factor $\xi$ as in the \tvc\ of
    the standard model in Fig. \ref{fig31e}.  Also shown in the figure
    are the observed terminal velocities in the bulge region. The
    cusped orbit and the other x$_1$-orbits at larger galactocentric
    radius represent the observed terminal velocities well. The
    velocities of x$_2$-orbits peak at $|v|\simeq 85 {\rm km/s}$ in
    the plot. }
  \label{fig35c}
\end{figure}

\subsection{Orbits and gas flow in the bulge region}
\label{s31}

In this section we consider in more detail the gas flow and \tvc\ in
the central $|l|\lta 10\deg$. The discussion of the large-scale
morphology and pattern speed is continued in \S\ref{s4}.

We begin with an analysis of closed orbits in the standard potential,
including a central cusp in the mass model as described in
Section \ref{s22}, and assuming the same values for the pattern speeds
as in the standard gas model.  The closed orbits are found with a
simple shooting algorithm. Despite the intrinsic time-dependence of
the potential in the bar frame, the closed orbits in the inner kpc
remain essentially unperturbed.  Figure~\ref{fig35b} shows closed
x$_2$- and x$_1$-orbits around the so-called "cusped" orbit, the
x$_1$-orbit whose turning points on the bar's major axis have a cusp
shape.  Closer to the galactic centre the more tightly bound
x$_1$-orbits become self-intersecting, and the x$_2$-orbit family of
stable orbits elongated perpendicular to the bar appears.  Note that
in this potential the outermost x$_2$ orbits are (nearly) converging
on the major axis of the bar, implying that gas clouds on these orbits
would collide.  This limits the radial extent of {\it accessible}
x$_2$-orbits for gas clouds in a hydrodynamic flow.

Between the cusped x$_1$ orbit and the first non-intersecting x$_2$
orbit there is a region in which no closed orbits suitable for gas
flow exist (Fig.~\ref{fig35b}).  Because of this, inflowing gas has to
quickly pass this region: the mechanism described by Binney \etal\ 
\shortcite{BGSBU}, where gas moving in from the last $x_1$-orbits
collides with gas on the outermost $x_2$-orbits, producing a spray
that then forms an off-axis shock by hitting the far side of the
$x_1$-orbits, cannot work as well in the potential here because of the
lack of suitable outer $x_2$-orbits. Instead, the main place of
dissipation of kinetic energy is likely to be the self-crossing loops
of the $x_1$-orbits inside the cusped orbit, from where the gas moves
inwards to hit the $x_2$-disk further in. This may explain why there
is a mostly gas-free gap in the hydrodynamic simulations between the
last non-intersecting x$_1$ orbit and the first accessible x$_2$
orbit, without any clear off-axis shocks like those found in other
barred galaxy models (e.g.  Athanassoula \shortcite{Athana92b}). 

How would gas clouds following the closed orbits in our standard model
potential compare to the terminal velocity observations?
Figure~\ref{fig35c}, plotted for bar angle $20\deg$, shows that the
terminal velocities of the closed x$_1$-orbits reproduce the
observations surprisingly well. The cusped orbit in Fig.~\ref{fig35c}
is not only at the same longitude as the maximum in the observed \tvc\ 
(at positive $l$ where we have data), but also appears to account for
the decline of the observed terminal velocities at lower $l$.  This is
consistent with a gas-free gap between the cusped orbit and the first
acceptable x$_2$-orbit, due to which no strong leading shocks form
like those observed in other barred galaxies.  Episodic infall of gas
clouds within the gap region may nonetheless form transient shocks
similar to those observed by H\"uttemeister et al.\ \cite{Hue98}.

The {\it closed orbits} thus reproduce the observed terminal
velocities, but why is there no gas at these velocities in the
simulations? Possible explanations are as follows:
First, the resolution length of our SPH
code, which is dominated by the smoothing length of the SPH particles,
may be too large to follow the strongly elongated x$_1$-orbits near
their cusped ends. Also, in the low density region further in, the
relative velocities of neighbouring particles are large, so that the
corresponding viscosity may lead to fast infall of SPH particles to
the centre. This would depopulate the inner x$_1$ orbits, where the
largest terminal velocities are expected. Indeed, the terminal
velocity curve attains larger peak velocities in high-resolution
models with some $10^5$ particles (the standard gas model has $\approx
6\cdot 10^4$ particles), for example the maximum is at $\approx 222 {\rm
  km/s}$ on the $l>0\deg$ side, compared to $\approx 210 {\rm km/s}$
in the standard gas model.

Second, the detailed orbit shapes in the inner kpc of the MW depend on
the fine-structure of the potential there, which is not accurately
known, both because of the limited resolution of the underlying NIR
data and luminosity model, and the difficulty of the deprojection in
this region. The true x$_1$ orbits could easily be somewhat less cuspy
near their ends, or the $x_2$-orbits less converging, making them more
easily populated with gas clouds.

Third, the hydrodynamic flow in the central MW could be genuinely
slower than suggested by the closed orbits which it approximately
follows.  Increasing the model terminal velocities in the bulge region
to match the observations would then require a somewhat larger bulge
mass-to-light ratio than the value determined from the \tvc\ fit at
$10\deg < |l| <50\deg$. This might be plausible if the disk population
is somewhat younger than that of the bulge.  It is unlikely that the
$M/L$ ratio is significantly larger in the entire central $\kpc$ of
the MW because the peak velocitiy of gas on x$_2$-orbits in our
standard gas model is $85 {\rm km/s}$, which compares well to the
observed $\lta 80 {\rm km/s}$ for CS-cloud cores at $|l|\lta 0.7\deg$,
where the projected model x$_2$-orbits are located. However, the
potential near $R\approx 1\kpc$, which is most relevant for the x$_1$
orbits, is more sensitive to the upper bulge component than that in
the inner $100\pc$ where the x$_2$ orbits are. The required variations
of the bulge $M/L$ to give terminal velocities of $\gta 250 {\rm
  km/s}$ would seem consistent with the spread of K band mass-to-light
ratios for other bulges. E.g., from NIR surface brightness photometry
of early type spirals, Moriondo et al.\ \shortcite{mor98} obtain a
mean value and dispersion $(0.6\pm 0.2) (M_{\odot}/L_{\odot})_K$ [the
velocity scale of the standard gas model corresponds to $(M/L)_K
\approx 0.6 (M_{\odot}/L_{\odot})_K$].

In summary, there are uncertainties in modelling
the gas velocities in the inner few hundred parsec of the
bulge. We will not pursue this
further here, because as already discussed in the last subsection,
this is not important for the gas flow in the main spiral
arms further out. Rather, we now turn to the determination of
the pattern speeds in the Milky Way. 

\begin{table*}
\begin{centering}
\begin{small}
\begin{tabular}{l|l|cc|l}
\hline
Model & Potential & $\Omega_p$ & $\Omega_{sp}$ & Remarks \\
\hline
20 & standard & $60 \Gyr^{-1}$ & $19.6 \Gyr^{-1}$ & Standard model \\
40 & standard & $61.4 \Gyr^{-1}$ & $40.8 \Gyr^{-1}$ &  \\
60 & standard & $61.4 \Gyr^{-1}$ & $61.4 \Gyr^{-1}$ &  \\
\hline
halo & standard & $60 \Gyr^{-1}$ & $19.6 \Gyr^{-1}$ & Standard model  \\
 & & & & including halo \\
\hline
bar50 & standard & $50 \Gyr^{-1}$ & $20 \Gyr^{-1}$ & \\
bar70 & standard & $70 \Gyr^{-1}$ & $20 \Gyr^{-1}$ & \\
\hline
open2 & two-armed, pitch angle similar  & $60 \Gyr^{-1}$ & $20 \Gyr^{-1}$ &  \\
 & to standard four-armed model & & & \\
2spi & two-armed, pitch angle half & 
$60 \Gyr^{-1}$ & $20 \Gyr^{-1}$ &  \\
 &  the value in the standard four-armed model & & & \\
mix & similar to four-armed model,  but Sag-Car arm & $60 \Gyr^{-1}$ & $20 \Gyr^{-1}$ & \\
&  and counter-arm are reduced in
 amplitude & & & \\
noarms & standard with spiral perturbation switched off & $60 \Gyr^{-1}$ & $20 \Gyr^{-1}$ & \\
\hline
incl10 & four-armed model, bar angle $10\deg$ & $60 \Gyr^{-1}$ & $20 \Gyr^{-1}$ & \\
incl15 & four-armed model, bar angle $15\deg$ & $60 \Gyr^{-1}$ & $20 \Gyr^{-1}$ & \\
incl25 & four-armed model, bar angle $25\deg$ & $60 \Gyr^{-1}$ & $20 \Gyr^{-1}$ &  \\
incl30 & four-armed model, bar angle $30\deg$ & $60 \Gyr^{-1}$ & $20 \Gyr^{-1}$ &  \\
\hline
strongarms & standard & $60 \Gyr^{-1}$ & $20 \Gyr^{-1}$ & $m\geq2$-multipoles  of density\\
& & & & outside $3.5\kpc$ multiplied by $1.5$ \\
tumblingbar & & $60 \Gyr^{-1}$ & $20 \Gyr^{-1}$ & standard potential, but the centre \\
& & & & of the bar perturbation is not in MW centre \\ 
\hline
\label{table1}
\end{tabular}
\caption{The gas models discussed in this paper.}
\end{small}
\end{centering}
\end{table*}

\section{The gas flow in models with separate bar and 
         spiral arm pattern speeds}
\label{s4}
\label{multipat}
\begin{figure*}
  \caption[]{The gas distribution of the standard model {\it 20} for
a sequence of evolutionary times and corresponding phase differences
between the bar and spiral components of the potential: $(0.296\Gyr,
318\deg), (0.304\Gyr ,337\deg), (0.312\Gyr, 356\deg)$ (upper row from
left to right), and $( 0.320\Gyr, 14\deg), (0.328\Gyr, 33\deg)$, and
$(0.336\Gyr, 51\deg)$ (lower row from left to right). The long axis
of the bar is aligned with the x-axis in all panels. Note the 
evolution in the connecting region between the inner and outer arms.}
  \label{seq32}
\end{figure*}  

Figure \ref{seq32} shows a time sequence for the gas distribution in
our point-symmetric standard model with pattern speeds $\Omega_p=60
\Gyr^{-1}, \Omega_{sp}=20 \Gyr^{-1}$. The inner arms emanating from
the ends of the bar corotate with the bar pattern speed; they are
clearly driven by the bar.  These inner arms are connected to the
outer spiral arms by a time-dependent transition region near bar
corotation.  Here a lateral arm from the distant end of the bar as
seen from the Sun merges with the other inner arm from the nearer end
in some frames (4-5), but not in others (1-2).  In the latter case,
the lateral arm continues into the outer spiral arm passing close to
the Sun, in the former both arms join into a weaker outer arm staying
well inside the Sun. The outer arms themselves move with respect to
the bar frame.  However, they do not rotate in the plot steadily
around the model's centre, as one might have expected if they were
driven by the different pattern speed of the spiral arm {\sl
  potential}, but apparently exhibit complicated back-and-forth
oscillations, with respect to each other and with respect to the bar
frame, and some arms merge and bifurcate at certain times; compare the
vicinity of $(x,y)=(-5,-3)$ in the different panels.

The outer spiral structure evolves in such a complicated way because
both the spiral arm potential and the bar simultaneously force the gas
distribution with different pattern speeds. To investigate this
further, we have computed a gas model similar to the standard model,
but with spiral structure removed from the gravitational potential, so
that the non-axisymmetric component of the potential is solely due to
the bar. The resulting model has four spiral arms inside the bar
corotation radius, and two symmetric outer spiral arms, all stationary
in the bar frame. In the model with a driving spiral structure
potential, the four outer arms can be regarded as a superposition of
one component generated by the bar perturbation, which is at nearly
constant position in the bar frame, and a second component driven by
the spiral structure potential perturbation. The former is the
stronger pair of arms in Fig.~\ref{seq32}, one of which passes just
inside the Sun symbol in the figure. The second component, visible as
the weaker pair of arms between the bar-driven arms in
Fig.~\ref{seq32}, can be seen to fall behind the bar-driven arms
through the sequence of frames in Fig.~\ref{seq32}
($\Omega_p\gt\Omega_{sp}$).  At certain times the second component is
seen to branch off the bar-driven arms (frame 1). Thereafter, the
spiral-driven arm appears to fall behind, move inwards, until it
finally collides with the opposite bar-driven arm (frames 2 through 6
$\simeq$ frame 1 through frame 3).  In the course of this evolution,
the detailed morphology of the transition region around bar corotation
changes.  At most times, the spiral-driven arms are connected to the
inner lateral arms (frames 2-5), at others the latter connect to the
bar-driven arms (frame 1). At certain times, an arm may look
fragmented, and its tangent point may split in longitude.

Thus when neither the bar nor the spiral structure component dominates
the non-axisymmetric potential and gas flow in the disk, as in
Fig.~\ref{seq32}, it is difficult to deduce from a single snapshot of
the arm morphology that the system supports a second, independent
pattern speed. This may explain why it is difficult to say from the
observed arm morphologies in barred spiral galaxies whether the spiral
arms are driven by the bar or not (see Sellwood \& Wilkinson
\shortcite{SellWil93}).

\subsection{Pattern speeds in the Milky Way}
\label{s41}

We now investigate Galactic models with different combinations of the
bar pattern speed $\Omega_p$ and the spiral arm pattern speed
$\Omega_{sp}$. Particularly for the spiral arms the assumption of a
constant pattern speed is probably still idealized, but represents a
first step towards understanding realistic cases.  In these models,
the separation of bar and spiral arm components in the potential is
based on the density distribution. We assume that the
$m\geq2$-multipoles of the density at galactocentric radii
$r<r_{\rm cut}=3.5\kpc$ (bar) rotate with $\Omega_p$, and outside of
$r_{\rm cut}$ (spirals) with $\Omega_{sp}$. The specific value of
$r_{\rm cut}$ was chosen because it corresponds approximately to the
end of the bar in the reference luminosity model of Paper II, and is
also equal to the inner radius of the spiral pattern there. All models
described in this Section are based on this luminosity model
(cf.~\S\ref{s21}).

Multiple pattern speeds ($\Omega_p\neq\Omega_{sp}$) imply a genuinely
time-dependent potential in the frame corotating with the bar.  For
each model we have therefore investigated a sequence of snapshots at
different evolutionary ages, separated by $\Delta\phi\approx 23\deg$
in phase difference between the two components in the potential.  This
corresponds to steps in the evolutionary age of the model of $\approx
0.01-0.02 \Gyr$, depending on the combination of pattern speeds. A
finer analysis of our standard gas model in steps of
$\Delta\phi\approx 9\deg$ showed that no significant features in the
models are missed with $\Delta\phi\approx23\deg$.  Because the
potentials used in this section are point-symmetric, we need to cover
only a range of $180\deg$ in bar-spiral phase difference.  All
analysed snapshots have evolutionary ages at or around $0.30 \Gyr$,
the time after which the gas flow in the similar single pattern speed
models of Paper I had become approximately quasi-stationary.

\begin{figure*}
  \caption[]{Gas distribution in models {\it 60} (left), {\it 20}
(middle), and {\it 40} (right).  Overplotted black dots indicate particles
highlighted in the \lvplot s in Fig.~\ref{lvplot32}.  The position of
the Sun at $(x,y)\approx(7.5\kpc,2.7\kpc)$ is shown by the $\odot$
symbol.}
  \label{gas32}
\end{figure*}

\begin{figure}
  \caption[]{Terminal velocity curves for the best evolutionary ages
of the single and double pattern speed models {\it 60}, {\it 20}, and
{\it 40}.  The model \tvcs\ are scaled by different factors $\xi$,
determined respectively from the best fit to the observations for
$10\deg\leq|l|\leq50\deg$. For clarity, the \tvcs\ are separated by
$40{\rm km/s}$ from each other in velocity.  }
   \label{tvc32}
\end{figure}

We begin with the three models {\it 60}, {\it 40} and {\it 20} with
different spiral arm pattern speeds $\Omega_{sp}=61.4, 40$ and $19.6
\Gyr^{-1}$.  The bar pattern speed is set to $61.4 \Gyr^{-1}$ ($60
\Gyr^{-1}$ for model {\it 20}), based on the results of Paper I.
Model {\it 60} is the single pattern speed model; the two other values
for $\Omega_{sp}$ are motivated as follows.  Firstly, Amaral \&
L\'epine \cite{AmLep97} preferred the lower value $\Omega_{sp}=20
\Gyr^{-1}$, based on the positions and ages of open clusters, and from
comparing their model with other data. Fern\'andez et al.\ 
\cite{Fer01} obtained a somewhat higher $\Omega_{sp}\approx 30 {\rm
  km/s/kpc}$ from Hipparcos data for OB stars and Cepheids,
supplemented with radial velocities or distances, respectively.  Thus
we selected $\Omega_{sp}=40 \Gyr^{-1}$ as an additional intermediate
value between $\Omega_{sp}=20 \Gyr^{-1}$ and the single pattern speed
model.  Secondly, for $\Omega_{sp}=60, 20, 40\, \Gyr^{-1}$, the bar
corotation radius coincides approximately with the corotation, inner
Lindblad, and inner ultra-harmonic (1:4) resonance of the spiral
pattern, respectively (see Fig.~\ref{figres}).

What are the characteristic differences between these models?  In
Figure~\ref{gas32} we compare their spatial gas distributions.  All
three models have four inner spiral arms inside bar corotation (at
$\approx3.4\kpc$), and four outer spiral arms outside of a transition
region beyond the end of the bar.  Because we have always selected
that snapshot of a model which best reproduces the observed \lvplot\ 
(Fig.~\ref{dameco}), the spiral arm tangent points are fixed relative
to the position of the Sun.  Thus in the figure the outer arms of all
models appear approximately at the same positions.

However, near the bar corotation radius the models differ
significantly. In the single pattern speed model {\it 60}, where the
spiral structure corotation radius coincides with that of the bar, the
gas in the corotation region moves with near-sonic velocities relative
to the pattern.  Consequently shocks are weak or nonexistent in this
region, and the spiral arms in the gas response dissolve there.  On
the other hand, for model {\it 40} with $\Omega_{sp}=40 \Gyr^{-1}$,
the corotation radius of the spiral structure is at $\approx5\kpc$.
Again, the gas distribution shows gaps in the spiral arms at this
location -- one pair of arms nearly vanishes there, the other pair
weakens -- but there is no such gap in the spiral arms near bar
corotation. Rather, the connection between the inner and outer spiral
arms is dynamic, due to the different pattern speeds.  Finally, in
model {\it 20} the spiral structure corotation radius occurs beyond
the solar orbit. In this model we see no clear gaps in the arms.
However, near the bar corotation radius $R_{\rm cr}^{\rm bar}\approx
3.4\kpc$, which coincides with the inner Lindblad resonance of the
spiral pattern, the arms weaken, and the transition region appears to
be more complicated than in model {\it 40}.

\begin{figure*}
\caption[]{\lvplot s for the best evolutionary times of models {\it
    60} (at age $0.30 \Gyr$, top), {\it 20} (at $0.32 \Gyr$, middle),
  and {\it 40} (also at $0.32 \Gyr$, bottom).  The model velocities
  are scaled by the same factor $\xi$ as the respective \tvc\ in
  Fig.~\ref{tvc32}; the LSR-velocity is $220 {\rm km/s}$.  Particles
  in low surface density regions are suppressed to enhance the
  contrast as described in the caption of Fig.~\ref{fig31g}. For
  comparison with observations, spiral arm ridge lines from
  Fig.~\ref{dameco}, positions of giant molecular clouds (``x''
  symbols) and HII regions (``+'' symbols) from Fig.~\ref{fig31g}, and
  spiral arm tangent directions from Englmaier \& Gerhard \cite{EG99}
  are overplotted.}
   \label{lvplot32} 
\end{figure*}

The comparison with the Galactic \tvc\ is shown in Figure~\ref{tvc32}.
All three models reproduce the observed data quite well, but the slope
in the region $10\deg\leq|l|\leq50\deg$ is fit significantly better by
the multiple pattern speed models than by the single pattern speed
model.  For $|l|\gta40\deg-50\deg$ the model \tvc s start to fall
below the observed \tvc; as we have discussed in \S\ref{s3}, for the
assumed $V_0=220\kms$ the dark halo starts to contribute to the
observed velocities there, according to our NIR-based models.

Figure \ref{lvplot32} shows \lvplots\ for the three models. To
facilitate their interpretation, we have overplotted a number of
observed features, and have highlighted particles belonging to
specific arm features both in this figure and in the density plots of
Fig.~\ref{gas32}, labelled by "A" and "B".  Feature ``A'' corresponds
to the 3kpc-arm in the observed CO \lvplot\ (Fig.~\ref{dameco}),
feature ``B'' to its symmetric counter-arm in the models (see
Fig.~\ref{gas32}; in model {\sl 20} the neighbouring arm ``A' ''
is included).

Particularly important is the existence of certain regions in the
\lvplot, e.g., next to the 3kpc arm (feature "A"), where hardly any
gas is found in the multiple pattern speed models.  Such voids,
designated by ``V'' in the plots, arise because the gas is aligned
morphologically and kinematically with spiral arms there, which appear
as well-defined ridges with adjacent voids in the \lvplot.  From
Fig.~\ref{dameco} we see that similar voids are also visible in the
observed $^{12}$CO \lvplot, specifically next to features ``A'' and
``B'' in the bar corotation region.  This suggests strongly that in
the MW the gaseous arms go through bar corotation.  On the contrary,
in the single pattern speed model {\it 60}, the arms dissolve in the
bar corotation region, the 3kpc-arm is thus imcomplete, and the gas is
spread out approximately evenly over the corresponding parts of the
\lvplot.  Models with a separate second pattern speed for the spiral
arms in the MW are therefore preferred over single pattern speed
models. Models with a growing bar amplitude also
support spiral arms in the corotation region (Thielheim \& Wolff
\shortcite{TW82}); however, it is likely that when self-gravity is
included and the amplitude becomes non-linear, the growing spiral
pattern will again develop an independent pattern speed.

Overall, the models in Fig.~\ref{gas32} provide a good match to the
observed CO \lvplot. The main features that are reasonably well
represented are: the arm tangent at $l=30\deg$; the observed
$l=25\deg$ tangent, although at smaller $l\simeq20\deg$ in models {\it
  20} and {\it 40}; the morphology of this arm; the location
of the main spiral arm leading to the $l=50\deg$ tangent; the
morphology of the ridges and voids at $-10\deg>l>-25\deg$ (apart from
model {\it 60}); the arm morphology around $(l,v)=(20\deg,60\kms)$.
The main weaknesses of the models, if we disregard the poorly resolved
bulge region, are: the arm in the models which, returning from the
$l=20\deg$ tangent, crosses the region $5\deg\lta l\lta10\deg$ at
$v>50{\rm km/s}$ and does not have a counterpart in the data, except
perhaps if shifted to lower velocities; the missing tangent at
$l=-30\deg$ in models {\it 60} and {\it 20}; the displacement of the
3kpc-arm towards lower velocities.

\begin{figure*}
  \caption[]{\lvplots\  for the best evolutionary time $0.307 \Gyr$ 
of model {\it bar50} (top, $\Omega_p=50 \Gyr^{-1}$), and for 
$0.321 \Gyr$ of model {\it bar70} (bottom, $\Omega_p=70 \Gyr^{-1}$),
with model velocities scaled
as described in the caption of Fig.~\ref{fig31e}. Overplotted data
are the same as in Figs.~\ref{fig31g} and \ref{lvplot32}.}
  \label{lvplot5070}
\end{figure*}   

In the model {\it 20}, the spiral arm corresponding to the Centaurus 
tangent at $l\approx-51\deg$ is strongest when one of the spiral-driven 
arms coincides with the bar driven-arm (e.g., frames 3-4 in
Fig.~\ref{seq32}). Shortly before and after this evolutionary time
this arm looks fragmented in the model, and its tangent point appears
split in longitude.  Indeed, there are observational indications that
the Centaurus tangent is split into two parts at $l\approx -50\deg$
and $\approx -55\deg-(-58\deg)$ (cf.\ Table 1 of Paper I).

The displacement of the 3kpc-arm in the model and the apparent absence
of its counter-arm in the data might indicate a non-point-symmetric
mass distribution in this region.  This possibility is discussed
further in Section \ref{s53}.  Also, in models where the spiral
pattern speed differs from the bar pattern speed, we expect also the
{\sl mass distribution} in the transition region to be generally
time-dependent. Thus in this region the mass distribution based on the
NIR data can represent only one snapshot in time. In addition, there
is not much information in the NIR data on the spiral arm heads in
this crucial transition region, to constrain the luminosity model of
Paper II. Hence the potential in this region is likely to be at best
approximately correct.

In summary, the multiple pattern speed models reproduce the observed
features in the bar corotation region better than the single pattern
speed model, in particular the regions void of gas in the \lvplot
near the 3 kpc arm.
Comparing the lv-plots for models {\it 20} and {\it 40}, we have found
a slight preference for model {\it 20} from the positions of the
spiral arms; however, the differences between these two models are too
small to determine the spiral arm pattern speed reliably.  We take the
model with $\Omega_{sp}\simeq 20 \Gyr^{-1}$ as our standard model, and
use this pattern speed in the following.

We now proceed to determine the best bar pattern speed $\Omega_p$.  To
this end we consider two models {\it bar50} and {\it bar70} which have
$\Omega_p=50 \Gyr^{-1}$ and $70 \Gyr^{-1}$, respectively.  For both
models, we set $\Omega_{sp}=20 \Gyr^{-1}$; since the differences
between models {\it 20} and {\it 40} are small, the precise choice of
$\Omega_{sp}$ should not matter.  Figure~\ref{lvplot5070} shows
\lvplots\ for the respective best evolutionary times of models {\it
  bar50} and {\it bar70}.  For both models the fit of the spiral
ridges in the inner disk region of the \lvplot\ is significantly worse
than for the models with $\Omega_p \approx60 \Gyr^{-1}$.  In model
{\it bar50} the positions of the spiral arm ridges and tangents are
worse (e.g., one of the $l>0$ tangents is missing), and the $l<0$ TVC
and the 3kpc-arm are particularly badly represented.  In model {\it
  bar70} the spiral arm ridge with tangent position at
$l\approx25\deg$ is at significantly too small longitudes, and the
same is true for the Centaurus tangent near $l\approx-51\deg$, which
corresponds to one of the bar-driven outer arms (see discussion
above).  It is noteworthy that the position and shape of the Centaurus
tangent are sensitive to $\Omega_p$ despite a galactocentric radius of
this tangent point of more than $6\kpc$.  Note also that a pattern
speed of $70 \Gyr^{-1}$ would put corotation inside the end of the NIR
bar (see \S\ref{s2} and Paper II).

\begin{figure}
  \caption[]{\tvcs\ at the respective best evolutionary time,
    for a sequence of models with bar angles $\phibar=10\deg$ (model
    {\it incl10}, at age $0.32 \Gyr$), $\phibar=15\deg$ (model {\it
      incl15}, at age $0.32 \Gyr$), $\phibar=25\deg$ (model {\it
      incl25}, at age $0.31 \Gyr$), and $\phibar=30\deg$ (model {\it
      incl30}, at age $0.30 \Gyr$).  For comparison, the standard
    model {\it 20} for bar angle $\phibar=20\deg$ is included in the
    figure.  In all models the pattern speeds are $\Omega_p=60
    \Gyr^{-1}$ and $\Omega_{sp}=20 \Gyr^{-1}$, and model velocities
    have been scaled by factors $\xi$, determined for each model by
    fitting the observed terminal velocities for $10\deg \leq |l| \leq
    50\deg$. For clarity, these \tvc s have been offset in steps of
    $40{\rm km/s}$, with model {\it incl15} plottet at the correct
    velocities.}
\label{incltvc}
\end{figure}   

We conclude that the best value for the bar pattern speed in the Milky Way
is $\Omega_p=(60\pm 5) \Gyr^{-1}$. For the preferred scaling of the
employed NIR bar model this corresponds to bar corotation at 
$3.4\pm0.3 \kpc$, equal to the length of the bar within
the uncertainties. I.e., the Milky Way bar is a fast bar.

\section{Bar orientation and influence of stellar spiral arms}
\label{s5}
\subsection{Bar angle}
\label{s51}

In Paper II, luminosity models were generated from the {\it
  COBE/DIRBE} L-band data for bar angles $\phibar=10\deg, 15\deg,
20\deg, 25\deg, 30\deg$ and $44\deg$ in the same way as for the
standard $\phibar=20\deg$ model. From considering the photometric
residuals and the line-of-sight distributions of clump giant stars in
the bulge, a preferred range for the bar angle
$15\deg\leq\phibar\leq30\deg$ was found. Here we obtain independent
constraints on $\phibar$ from corresponding gas flow models.

\begin{figure*}
  \caption[]{\lvplots\ for the gas flow models with bar angles $10\deg$ 
    (top left), $15\deg$ (top right), $25\deg$ (bottom left) and
    $30\deg$ (bottom right), for the same evolutionary times as in
    Fig.~\ref{incltvc}.}
  \label{incllv}
\end{figure*}   

\begin{figure}
  \caption[]{\tvcs\  at the respective best evolutionary time,
    for gas models forced by different types of spiral structure in
    the mass density.  Model {\it open2} is shown at age $0.30 \Gyr$,
    model {\it 2spi} at age $0.30 \Gyr$, and model {\it mix} at age
    $0.31 \Gyr$, model {\it strongarms} at age $0.32 \Gyr$.  Also
    shown is the standard model {\it 20}, and model {\it noarms}
    without any massive spiral arms.  For all models, the pattern
    speeds are $\Omega_p=60 \Gyr^{-1}$ and $\Omega_{sp}=20 \Gyr^{-1}$,
    and the velocity scale is fixed by fitting to the observed
    terminal velocities for $10\deg \leq |l| \leq 50\deg$. For
    clarity, the model \tvcs\ are offset in steps of $40{\rm km/s}$,
    with model {\it open2} plotted at the correct velocities.  }
\label{tvcspi}
\end{figure}   

For each value of the bar angle, the luminosity model was converted to
a mass model assuming constant M/L. As described above, the
non-axisymmetric part of the potential was split into bar and spiral
arm components, and gas models were computed with pattern speeds
$\Omega_p=60 \Gyr^{-1}$ and $\Omega_{sp}=20 \Gyr^{-1}$, respectively.
In Figures \ref{incltvc} and \ref{incllv} we show the \tvc s and
\lvplots\ of models {\it incl10}, {\it incl15}, {\it incl25} and {\it
  incl30} with bar angles $\phibar=10\deg, 15\deg, 25\deg, 30\deg$ at
their respective best evolutionary times, as judged by comparing their
\lvplot s with Fig.~\ref{dameco}.

Model {\it incl25} reproduces the terminal velocity observations with
similar quality as the standard gas model {\it 20}, but model {\it
  incl30} is clearly inferior -- the terminal velocities are too large
at positive longitudes and too small in modulus at negative
longitudes.  Obviously, this cannot be corrected for by a change of
the velocity scaling factor $\xi$. The \tvcs\ of models {\it incl10}
and {\it incl15} are also inferior fits to the observed terminal
velocities, albeit not as bad as model {\it incl30}.

The \lvplots\ of these models depend strongly on the assumed bar
angle. Models {\it incl10}, {\it incl15} and {\it incl30} do not
reproduce the observed spiral arm ridges. Models {\it incl10} and {\it
  incl15} also do not have a well-define 3 kpc arm at positive
longitudes, and model {\it incl30} shows very little arm structure at
all in the region around bar corotation.  Model {\it incl25}
reproduces the observations for a large range of longitudes similarly
well as the standard model, but the positions of the spiral arm ridges
at $-30\deg\lta l \lta-10\deg$ are not matched well and the
non-circular velocities in the 3 kpc arm are smaller than in the
standard $\phibar=20\deg$ model.  From the \lvplots\ model {\it 20} is
best, but we consider model {\it 25} as still satisfactory.

One may ask, how much of the difference between these models is due to
the different shapes of their underlying luminosity distributions and
gravitational potentials, and how much of it is due simply to the
different viewing geometries with respect to the gas flow? To answer
this question, we have constructed the \lvplot\ of the standard gas
model {\it 20} seen from a viewing angle of $\phibar=30\deg$. The
morphology of the arms in this \lvplot\ differs little from that in
the original $\phibar=20\deg$ \lvplot\ because of the tightly wound
pattern. Thus the differences between the \lvplot s in
Fig.~\ref{incllv} must be caused mainly by genuine changes in the gas
flow, originating from the different mass distributions corresponding
to the COBE data for different $\phibar$.

\subsection{Spiral arm models}
\label{s52}

\begin{figure*}
  \caption[]{\lvplots\ of the models with
    \tvcs\ shown in Fig.~\ref{tvcspi}, with their respective velocity
    factors $\xi$.  From top to bottom: model {\it open2}, model {\it
      2spi}, and model {\it mix}.}
  \label{lvspi}
\end{figure*}

The MW very probably has four spiral arms in the gas distribution
(e.g., Englmaier \& Gerhard \shortcite{EG99}).  It is not clear, however, whether
these also correspond to four {\sl stellar} spiral arms, because some
of the tangent points are not clearly seen in the near-IR light (see
the discussion in the Introduction and in Drimmel \& Spergel
\cite{Drimmel01}).  Here we investigate the \lvplot s that result when
different stellar spiral arm patterns drive the gas flow, and find
evidence for a four-armed spiral pattern also in the distribution of
luminous mass. Specifically, we have studied three models:
\begin{description}
\item[Open2:] A two-armed model with the same spiral arm pitch angle
  $13.8 \deg$
  as in the standard (four-armed) mass model, but where the Sag-Car
  arm and its counter-arm are removed from the model. Such a model can
  be justified by the fact that the Sag-Car arm is hardly visible in
  the NIR.
\item[2spi:] Another two-armed model, but with approximately half the
  pitch angle of the standard four-armed model. This model reproduces
  approximately the same tangent point positions on the sky as the
  standard model.
\item[Mix:] Similar to the standard mass model, but the Sag-Car arm
  and its counter-arm are given only $40\%$ of the peak density
  amplitude of the other two arms. This is based on the result of
  Drimmel \& Spergel \cite{Drimmel01}, who found that they had to
  reduce the amplitude of the Sag-Car arm in their best-fit four-armed
  model for the {\it COBE/DIRBE} J and K-band NIR maps.
\end{description}

Luminous mass models with these patterns were derived using the
algorithm described in Paper II, all for bar angle $\phibar=20\deg$,
by incorporating the respective spiral arm model in both the
parametric initial model and in a penalty term for the non-parametric
deprojection. As in the standard mass model the non-parametric
algorithm changes the spiral arms somewhat, but the overall pattern
stays intact.  Because the NIR data constrain only the arm tangent
points (Paper II), two-armed and four-armed models which reproduce the
tangent point data fit the photometry with similar quality.

\begin{figure*}
  \caption[]{Lv-plot for model {\it strongarms} that has 
    strong spiral arms in the potential, at its best evolutionary age
    $0.32 \Gyr$.  Model velocities have been scaled by an appropriate
    factor $\xi$, as described in the caption of Fig. \ref{incltvc}.}
  \label{fig34f}
\end{figure*}   

The \tvcs\ of gas flow models computed in the corresponding
gravitational potentials are shown in Figure~\ref{tvcspi}.  The
overall slope with $l$ of all these model \tvcs\ is similar, because
this is dominated by the monopole term in the mass distribution.
However, all three models {\it open2}, {\it 2spi} and {\it mix} do not
fit the wavy structure of the \tvc\ data as well as the standard model
{\it 20} (Fig.~\ref{fig31e}), with model {\it mix} the best among the
three. For comparison we also include in Fig.~\ref{tvcspi} the \tvc\ 
of a model {\it noarms}, whose gas flow was determined in a potential
that includes only the perturbation from the bar, but not that from
the spiral arms.  This model has only two outer spiral arms in the gas
distribution.

\lvplots\ of these models are shown in Figure~\ref{lvspi}.  Models
{\it open2} and {\it 2spi} compare poorly to the observed CO \lvplot:
the up-turning arm at $l\approx25\deg$ is missing, and the spiral
structure is generally wrong for $-15\deg\lta l\lta-35\deg$.  In both
these models the fit to the data is bad because the gas distribution
of the model is only two-armed. We have checked that this is not a
result of choosing the wrong bar and spiral arm pattern speeds, by
computing gas models in the two-armed potential of model {\it open2}
for the additional combinations of $(\Omega_p=50 \Gyr^{-1},
\Omega_{sp}=20 \Gyr^{-1})$ and $\Omega_p=\Omega_{sp}=60 \Gyr^{-1}$
(single pattern speed).  In both cases, the global, two-armed
morphology of the gas flow is the same as in model {\it open2}. Model
{\it mix} is the best of the three models in this section.  However,
the envelope of its lv-plot shows stronger bumps than the standard
model, in particular near $-30\deg\ldots -15\deg$.  The gravitional
potential of model {\it mix} is quite similar to our standard
potential, and therefore the similarity of the gas flows is expected. From 
these tests we conclude that a four-armed spiral arm potential 
is preferred (see also Englmaier \& Gerhard \shortcite{EG99}
and Fux \shortcite{Fux99}).

\begin{figure*}
  \caption[]{\lvplot\ for the best evolutionary age ($0.32 \Gyr$) 
    of model {\it tumblingbar}.  In this model the centre of the
    stellar bar is offset from the galaxy centre by $300\pc$, and
    rotates around it with $-\Omega_p$. Note that the 3kpc-arm is
    reproduced well, but not its counter-arm.}
  \label{tumblingbar}
\end{figure*}

\subsection{Stronger spiral arms}

The standard gas model fits well most spiral arms in the observations.
However, the famous 3kpc-arm, a prominent feature which extends from
$(l\approx 10\deg, v=0 {\rm km/s})$, through $(l=0\deg,v\approx -50
{\rm km/s})$, to $(l\approx -22\deg, v\approx -120 {\rm km/s})$ is
still displaced towards lower non-circular velocities. Because the
underlying mass model already contains spiral arms, assigning
gravitating mass to the gas particles does not lead to an improvement
(as a corresponding model confirmed).

However, compared to near-IR-observations of other spiral galaxies
(cf. Rix \& Zaritsky \shortcite{Rix95}), our standard mass model has rather
narrow and weak spiral arms. This might be simply because only the
positions of the spiral arm tangents are constrained by the {\it
  COBE/DIRBE} L-band data, while the spiral arm heads near the bar are
not well-constrained. It is therefore possible that our
standard mass model underestimates the strength of the Galactic spiral
arms. We have therefore investigated a model {\it strongarms} in which
we enlarged non-axisymmetric forces in the disk, by multiplying with
$1.5$ the $m\geq2$ multipoles of the spiral arm component (i.e., the
component corresponding to the density outside of $r_{\rm 
  cut}=3.5\kpc$) in the standard model, and computed the gas flow
in this modified potential.

We show in Figure~\ref{tvcspi} the \tvc\ and in Figure~\ref{fig34f}
the \lvplot\ of this model {\it strongarms}.  The fit to the \tvc\ 
observations is not as good as for the standard model {\it 20}, but in
the \lvplot\ the spiral arm ridges are quite similar.  Due to the
stronger spiral arm gravity, the bumps in the terminal velocity are
somewhat stronger, but still in the acceptable range.  Interestingly,
the 3kpc-arm now fits the observations nearly perfectly.  This shows
that the strength of the spiral arms is an important parameter for the
observed kinematics of the 3 kpc arm, and the non-circular motions in
this region of the Galactic \lvplot\ cannot simply be used to
determine the bar aspect angle (cf.\ Weiner \& Sellwood 1999).  It
also suggests that our standard model can be improved when a better
spiral arm model becomes available.

\subsection{Asymmetric Models}
\label{s53}
There are indications in the HI and CO surveys that the inner Galaxy's
gas distribution deviates significantly from point-symmetry with
respect to the centre. An example is the 3kpc-arm, which has no clear
counter-arm in the observed \lvplots. That is not to say that there is
no counter-arm; if asymmetric, its inner parts could for example
appear at similar locations in Fig.~\ref{dameco} as the arm which reaches
the \tvc\ at $l\approx 25\deg$.  All symmetric mass models, however,
yield counter-arms with about the same absolute velocities as the
3kpc-arm. A way out of this dilemma has been shown by Fux
\cite{Fux99}.  In his model, the 3kpc-arm and its counter-arm are
significantly disturbed by strong non-axisymmetric modes in the
Galactic centre, as well as in the outer disk. Such a mechanism may
also help to explain that the peak in the terminal velocity at
$l=+2\deg$ appears much higher than at $l=-2\deg$ in the CO data. In
addition, the x$_2$-disk in the centre may be disturbed by this
effect, and this might explain the uneven gas distribution seen in CS.
There is no evidence for an asymmetric mass distribution in the NIR
data, so in our mass models we can only introduce asymmetry in the
density by hand and study the consequences.

To this end, we first created uneven $m$ modes in the disk by
weakening one or two arms in the initial model, or by moving two
spiral arms closer together. In such gas models we observed strong
effects on the position of gas shocks and their relative strength in
the outer disk. Some cases look similar to the result of Fux with an
almost 3-armed outer disk structure. The inner arms, especially the
3kpc-arm, did not change much, however. Only in an extreme case have
we been able to move the 3kpc-arm to higher velocities, while
simultaneously the counter-arm was moved to lower velocities, but this
model does not fit the observations well overall.

In a second class of asymmetric models, we let the stellar bar centre
rotate on a circular orbit with radius $R_{\rm bar}$ and pattern speed
$\Omega_c\!=\!-60 \Gyr^{-1}\!=\!-\Omega_{p}$, i.e., the centre rotates
backwards with respect to the bar and with the same pattern speed as
the bar.  This introduces also a third parameter, the phase
$\alpha_{\rm bar}$ of the bar centre rotation at model age $0.00
\Gyr$.  This approach was motivated by $N$-body simulations
(Debattista, work in progress).  In simulation {\it tumblingbar}
($\Omega_c\!=\!-60 \Gyr^{-1}, R_{\rm bar}\!=\!800\pc, \alpha_{\rm bar}
\!\approx\! 80\deg$), we see the 3kpc-arm and its counter-arm move in
the right direction in the model's \lvplot\ (Figure~\ref{tumblingbar})
and the $3\kpc$-arm fits the observations well. However, we have not
found a snapshot of a model at which the 3kpc-arm, its counter-arm and
the overall spiral pattern all fit the data well. 

Obviously, the available freedom in introducing deviations from
point-symmetry is very large. We have only tested a few attractive
possibilities, and these models show that asymmetries in the Galactic
mass distribution may be important and could be at the root of some of
the remaining problems in our standard model.

\section{Summary and Conclusions}
\label{s7}
We have used new gas flow models to investigate the dynamics of the
Milky Way (MW) Galaxy from observed \lvplots.  Steady-state gas flows
in rotating, point-symmetric gravitational potentials for the Galactic
bar and disk were determined with SPH simulations. The potentials
were derived from non-parametric estimates of the spatial
near-infrared luminosity density (Bissantz \& Gerhard
\shortcite{BissGerh02}), based on the de-reddened {\it COBE/DIRBE} L-band
map of Spergel et al.\ \cite{Spergel96}, but also incorporating clump
giant star count data from Stanek \etal\ \shortcite{Stanek94},
\shortcite{Stanek97}.  The luminosity models contain a spiral arm model
for the disk, and were converted to mass models assuming constant
mass-to-light ratio in the inner MW.

Our best gas flow model gives a very good fit to the Galactic terminal
velocity curve for $|l|>15\deg$, and to the spiral arm ridges in the
observed CO \lvplot. This has enabled us to investigate a number of
dynamically important parameters such as the bar and spiral arm
pattern speeds, the multiplicity of the spiral structure in the
potential, and the bar angle. The main results from this study are as
follows.

1) In gas flow models with separate pattern speeds $\Omega_p$ for the
bulge/bar and $\Omega_{sp}$ for the spiral pattern, the spiral arms go
through the bar corotation region. Thus \lvplots\ of such models show
well-defined spiral arm shocks (ridges) through corotation, next to
areas which appear to be nearly void of gas.  By contrast, in single
pattern speed models the spiral arms dissolve in the bar corotation
region, so that the gas fills this region of the \lvplot\ 
approximately evenly, and no voids exist.

2) Similar voids are visible in the observed $^{12}$CO \lvplot. From 
a comparison with model \lvplots\ with different $\Omega_{sp}$
but similar $\Omega_p$ we find evidence for separate pattern speeds
in the Milky Way. The existence of self-consistent models with 
separate bar and
spiral arm pattern speeds was demonstrated by Rautiainen \& Salo
\shortcite{Rau99} in a study of two-dimensional N-body simulations,
some of which included a massless, dissipative gas component.
Models with a growing bar amplitude also
support spiral arms in the corotation region  (Thielheim \& Wolff
\shortcite{TW82}); however, it is likely that when self-gravity is
included and the spiral arm 
amplitude becomes non-linear, the growing spiral
pattern will again develop an independent pattern speed.

3) From a series of models the preferred range for the bar pattern
speed in the MW is $\Omega_p=60\pm 5 \Gyr^{-1}$, corresponding to
corotation at $3.4\pm 0.3 \kpc$.  This agrees well with
previous pattern speed determinations by Englmaier \& Gerhard
\cite{EG99}, Dehnen \cite{Dehn00}, and Debattista \etal\ \cite{Deba02}.
The bar pattern speed is well constrained because it influences not
only the inner spiral structure, but also the position of two outer
spiral arms in the lv-plot.  Models with $\Omega_p=50 \Gyr^{-1}$ and
$\Omega_p=70 \Gyr^{-1}$ are inferior.

The spiral arm pattern speed is less well constrained. Our preferred
value is $\Omega_{sp}\approx 20 \Gyr^{-1}$, but models with larger
$\Omega_{sp} \lt \Omega_p$ give only marginally inferior fits to the
observed \lvplot.

4) Gas flows in models which include massive spiral arms clearly fit
the observed $^{12}$CO ($l,v$) plot better than if the potential does
not include spiral structure. Furthermore, comparing models with two
and four arms in the gravitational potential, we found that only
models with four massive arms reproduce the Galactic \lvplot, while
gas flows in two-armed potentials do not resemble the spiral arm
pattern of the Milky Way.

In Galactic models with four-armed potentials and separate spiral arm
pattern speed, the gas flow has two pairs of inner arms which rotate
with the bar (lateral, and corresponding to the 3 kpc arm), and four
outer spiral arms which exhibit a complicated, time-dependent
back--and--forth oscillation in the bar frame. The outer and inner
spiral structures are connected by a time-dependent transition region
around bar corotation.

5) From a further series of gas models computed for different bar
angles, using separately determined luminosity models and
gravitational potentials as in Bissantz \& Gerhard
\shortcite{BissGerh02}, we found a range of acceptable bar angles
$20\deg\lta\phibar \lta25\deg$. The models for $\phibar= 15\deg$ and
$30\deg$ are clearly inferior, which is mainly due to differences in
the inferred gravitational potential.

The model with ($\phibar=20\deg$, $\Omega_p=60\Gyr^{-1}$,
$\Omega_{sp}=20 \Gyr^{-1}$) gives an excellent fit to the Galactic
terminal velocity curve for $10\deg \leq |l| \leq 50\deg$, and to the
gaps and spiral arm ridges in the observed CO \lvplot. There are still
discrepancies in the bar corotation region where the potential has
uncertainties: the 3 kpc arm has too low non-circular velocity, and
its counterarm is missing in the data. In the bulge region, closed
orbits reproduce the \tvc\ well, while the gas model has lower
velocities. This may be a resolution problem in the SPH model, but
could in part also be due to uncertainties in the potential which
influence the orbit shapes.

6) The 3 kpc arm non-circular velocities can be reproduced by a model
in which we artifically increased the $m\geq 2$-multipoles of the
spiral potential component by $1.5$ while keeping all other dynamical
parameters fixed. This is well within the uncertainties. Guided by a
number of asymmetries in the observed Milky Way gas distribution, we
also investigated potentials which are no longer point-symmetric.
Some of these models improved the fit to the 3kpc-arm and its
counter-arm. Although we did not find a model which at the same time
reproduces the entire \lvplot\ as well as our standard model, these
models suggest that such asymmetries may be important for better
understanding the gas flow in the inner Milky Way.

\section*{Acknowledgments}

This work was supported by grant 20-64856.01 of the Swiss
National Science Foundation.

\end{document}